\documentclass[conference]{IEEEtran}
\usepackage{amsmath}
\usepackage{amssymb}
\usepackage[dvips]{graphicx}
\usepackage{epsfig}

\newtheorem{prop}{Proposition}

\begin{document}


\title{Goodput Maximization in Cooperative Networks with ARQ}



%
\author{\authorblockN{Qing Chen and Mustafa Cenk Gursoy}
\authorblockA{Department of Electrical Engineering\\
University of Nebraska-Lincoln, Lincoln, NE 68588\\ Email:
chenqing@huskers.unl.edu, gursoy@engr.unl.edu}}


\maketitle

\begin{abstract}\footnote{This work was supported by the National Science Foundation under Grants CCF -- 0546384 (CAREER) and CNS--0834753.}
In this paper, the average successful throughput, i.e., $goodput$,
of a coded 3-node cooperative network is studied in a Rayleigh
fading environment. It is assumed that a simple automatic repeat
request (ARQ) technique is employed in the network so that
erroneously received codeword is retransmitted until successful
delivery. The relay is assumed to operate in either
amplify-and-forward (AF) or decode-and-forward (DF) mode. Under
these assumptions, retransmission mechanisms and protocols are
described, and the average time required to send information
successfully is determined. Subsequently, the goodput for both AF
and DF relaying is formulated. The tradeoffs and interactions
between the goodput, transmission rates, and relay location are
investigated and optimal strategies are identified.
\end{abstract}
\vspace{-.1cm}
\section{Introduction}
In wireless networks, automatic repeat request (ARQ) techniques have been applied to improve the
transmission reliability above the physical layer (PHY). Prior work (e.g., in
\cite{PN} and \cite{RPC}) has shown that in the PHY of coded or uncoded
systems, a higher transmission rate $R$ results in
higher packet error rates, leading to more ARQ retransmissions, while
a lower $R$ leads to reduced packet error rates and therefore less ARQ
retransmissions. Hence, choosing to transmit at very high rates can lead to low average successful throughput (i.e., goodput) due to increased number of retransmissions. On the other hand, transmitting at very low rates leads to more reliable communication but the goodput is also low due to low rates. Consequently, it is of significant interest to jointly optimize the transmission rates (at the physical layer) and the number of ARQ retransmissions (at the data-link layer) by adopting a cross-layer framework so that the goodput of the system is maximized.

In recent years, cooperative operation through relaying has attracted much interest due its promise to improve the link reliability \cite{CCN} . For instance, when the
source-destination link suffers severe fading, information can be sent to the destination through a relay node more reliably if the source-relay and relay-destination channels experience more favorable fading conditions. Note that such diversity achieved through cooperation can also lead to increased goodput. Hence, cooperation is another important tool that can improve the performance.

In this paper, we consider a 3-node cooperative network and investigate the maximization of the goodput in both amplify-and-forward (AF) and decode-and-forward (DF) relaying scenarios. In particular, we investigate the tradeoffs and interactions between the goodput, transmission rates, different ARQ mechanisms, different relaying schemes and relay locations. In both AF and DF modes, we first quantify the link
error probabilities through the capacity outage formulation in a coded
system in Rayleigh fading channels. Then, we analytically identify the average number of
transmissions required for successful delivery, and formulate the goodput of the system. Through numerical results, we study the tradeoffs between the different parameters of the network.

The remainder of this paper is organized as follows. Section II introduces the network
model and channel assumptions. Section III  presents the goodput
analysis in AF and DF cooperative networks with ARQ. Numerical
results are given in Section IV. Finally, Section V provides the
conclusions.


\vspace{-.2cm}
\section{System Formulation and Channel Assumptions}
We consider a 3-node cooperative
network in which the source sends information to the destination with the aid of the relay node.   We assume that the
source-destination (S-D), source-relay (S-R), and
relay-destination (R-D) links experience independent Rayleigh fading.

We have the
following key assumptions: 1) channel codes support communication at the instantaneous channel capacity levels, and outages, which occur if transmission rate exceeds the instantaneous channel capacity, lead to packet errors and are perfectly detected at the
receivers; 2) depending on whether packets are successfully received or not, ACK or NACK control frames are sent and overheard;
3) ACK/NACK is received reliably with no errors; 4) each
codeword contains a certain number of data packets and the transmission
of data packets begins when source broadcasts to both relay and
destination; 5) the relay mechanism is assumed to be incremental
relaying in which the relay node doesn't need to engage in transmission whenever
the S-D link is successful \cite{RHF},\cite{WDB}.

The ARQ process differs in AF and DF modes. In the AF mode,
destination sends ACK when packets are successfully received from
the source to notify the source to schedule the next codeword
transmission. Otherwise, if source-destination (S-D) link fails, an
NACK is sent and relay amplifies and forwards the packets to the
destination via the source-relay-destination (S-R-D) cooperative
link. If the transmission through the S-R-D link fails as well,
destination sends a second NACK and source retransmits the codeword.

In the
DF mode, similarly as in AF, it is initially checked whether the transmission through the S-D link is successful. If not, NACK is sent and it is checked whether the relay, which also has to decode the incoming packets and send ACK/NACK frames upon its correct/erronous reception of packets, has received the packets successfully through the source-relay (S-R) link. If the S-R link has also failed, retransmission from the source is requested. If, on the other hand, relay has decoded the packets correctly, relay forwards the packet to the destination. In cases of failure in the relay-destination (R-D) link, retransmissions are requested from the relay. Therefore, retransmissions can be confined only to the R-D link when
the relay already has the correct codeword. The network states will be further elaborated in Section III.

\subsection{Cooperative Network Model}
The introduction of a relay node definitely adds more energy
overhead and system complexity, but it creates more reliability from
the auxiliary R-D link when S-D channel fails. If we assume the
signal transmitted from the source is $x$ with unit energy, the received signals at the relay and destination can be
represented by
\vspace{-.3cm}
\begin{align}
y_{s,r}=\sqrt{P_{t}}h_{s,r}x+n_{s,r}\label{eq:ysr}
\\
y_{s,d}=\sqrt{P_{t}}h_{s,d}x+n_{s,d}\label{eq:ysd}
\end{align}
where $P_{t}$ is the transmit power of the source. $h_{s,r}$ and
$h_{s,d}$ are channel fading coefficients in the S-R and S-D links, respectively. The fading coefficients are assumed to be zero-mean circularly symmetric
Gaussian complex random variables, modeling Rayleigh fading. Path loss is also incorporated into the model through the variance of the fading coefficients. $n_{s,d}$ and $n_{s,r}$ are the
additive Gaussian noise components (with variance $N_0$) at the destination and relay. We define
the SNR at the transmitter side by $$\gamma=\frac{P_{t}}{N_{0}}.$$

\subsection{Goodput on a Single Link}
Suppose that the source has $J$ codewords to transmit to the destination and the
transmission rate is $R$ bits per second. Each codeword has $L$ bits
and Codeword(i) needs $X_{i}$ number of transmissions before being successfully
received at the destination. The long-term effective
throughput on this single link can be formulated by the total
transmitted bits over total transmission time \cite{PN}. In the limit as
$J\rightarrow\infty$, we have
\begin{align} \label{eq:goodput-singlechannel}
\eta=\lim_{J\rightarrow\infty}\frac{JL}{\Sigma_{i=1}^{J}(X_{i}\frac{L}{R})}=\lim_{J\rightarrow\infty}\frac{R}{\frac{1}{J}\Sigma_{i=1}^{J}X_{i}}=\frac{R}{E(X)}.
\end{align}
$X$ is a geometric random variable \cite{PN}, \cite{SBAM} and
$E(X)=\frac{1}{1-\varepsilon}$ where $\varepsilon$ is the link error
probability. The link error probability is defined as the probability associated
with the outage event on a specific link \cite{CCN}. For a Rayleigh
fading channel, the fading coefficient $h$ is modeled as a zero-mean
complex Gaussian random variable with variance $\sigma^2$. So
$|h|^2$ is exponential distributed with mean
$\frac{1}{\lambda}=\sigma^2$. Therefore, the link error probability
for given SNR $\gamma$ and transmission rate $R$  is
\begin{align}
\varepsilon=P(\log_{2}(1+|h|^2\gamma)<R)&=P(|h|^2<\frac{2^R-1}{\gamma}) \nonumber
\\
&=1-e^{(-\frac{2^R-1}{\gamma\sigma^2})}
\end{align}
\vspace{-.5cm}
\section{Goodput Analysis with ARQ}
\subsection{Goodput in AF Mode} 
\begin{figure}
\begin{center}
\includegraphics[width = 0.3\textwidth]{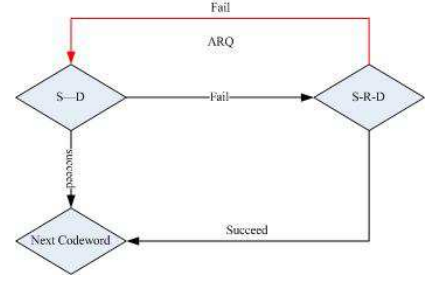}
\caption{Transmission Flow in AF} \label{fig:AF-routing}
\end{center}
\end{figure}

In the AF mode, relay amplifies and
forwards the packets to the destination only if the S-D link is in outage.
Hence, there are totally 3 mutually exclusive working states in AF:
1) S-D link is successful; 2) S-D link is in outage, but S-R-D link is
successful; 3) both S-D link and S-R-D link are in outage. The source
schedules the transmission of the next codeword only after the destination
sends back an ACK to announce that the codeword has been successfully
received through either State 1 or 2. In State 3,
destination sends back an NACK to have the source to retransmit the codeword, triggering ARQ. Consequently, a new transmission round is
started all over. This procedure is repeated until a successful codeword delivery is achieved
at the destination. The transmission flow is illustrated in Fig.
\ref{fig:AF-routing} where the red branch denotes the ARQ process.

We normalize the variance on the S-D link as $\sigma_{s,d}^2=1$. From (4), we know the outage probability on the S-D link is
\begin{align}
\varepsilon_{1}&=1-e^{-\frac{2^R-1}{\gamma}}.
\end{align}
The variances of the other links are assumed to be proportional to $\sigma^2 \varpropto d^{-\alpha}$ where $\alpha$ is the path loss coefficient \cite{RHF}. As for the S-R-D cooperative link, by using a generalized distance
factor $k=\frac{d_{s,r}}{d_{s,d}}, (k\in(0,1))$, the received SNR
via this link can be formulated as
\begin{align}
\gamma_{s,r,d}&=\frac{\gamma|h_{s,r}|^2\gamma|h_{r,d}|^2}{\gamma|h_{s,r}|^2+\gamma|h_{r,d}|^2+1}\nonumber\\
&=\frac{k^{-\alpha}\gamma|\hat{h}_1|^2(1-k)^{-\alpha}\gamma|\hat{h}_2|^2}{k^{-\alpha}\gamma|\hat{h}_1|^2+(1-k)^{-\alpha}\gamma|\hat{h}_2|^2+1}.
\end{align}
where $|\hat{h}_1|^2$ and $|\hat{h}_2|^2$ are independent random variables with the same distribution as $|h_{s,d}|^2$.
Note that we implicitly assume with the above formulation that the we have a linear 3-node network where the relay lies in between the source and destination. A more general scenario in which the relay has a certain vertical distance to the line that connects the source and destination can be treated by normalizing $k$ with the cosine of the angle between the source and relay.
From \cite{MO}, we know that $\gamma_{s,r,d}$ has a CDF given by
\begin{align}
p(\gamma_{s,r,d}<\gamma_{th})=1-\sqrt{\xi}K_{1}(\sqrt{\xi})e^{-\gamma_{th}\left(\frac{k^{\alpha}}{\gamma}+\frac{(1-k)^{\alpha}}{\gamma}\right)}
\end{align}
where
$\xi=\frac{4(\gamma_{th}^2+\gamma_{th})}{\gamma^2(k(1-k))^{-\alpha}}$
and $K_{1}()$ is the first order modified Bessel function of the
second type. Hence, the outage probability on the S-R-D link is
\begin{align}
\varepsilon_{2}=1-\sqrt{\xi}K_{1}\left(\sqrt{\xi}\right)e^{-(2^R-1)\left(\frac{k^{\alpha}}{\gamma}+\frac{(1-k)^{\alpha}}{\gamma}\right)}
\end{align}
where $\xi=\frac{4((2^R-1)^2+(2^R-1))}{\gamma^2(k(1-k))^{-\alpha}}$.
Now, the probabilities of the 3 network states of the AF mode are
\begin{align}
p_{1}&=1-\varepsilon_{1} \label{eq:p_1AF}\\
p_{2}&=\varepsilon_{1}(1-\varepsilon_{2}) \label{eq:p_2AF}\\
p_{3}&=\varepsilon_{1}\varepsilon_{2} \label{eq:p_3AF}.
\end{align}
We define $T=\frac{L}{R}$ as the time required to transmit one codeword
over any link. In State 1, a duration of $T$ is needed. At the end of this period, successful transmission is achieved. In both States 2 and
3, a duration of $2T$ is needed. In the first $T$ seconds, source transmits. Since the S-D link fails in these two states, the relay engages in transmission in the following $T$ seconds. As demonstrated in (\ref{eq:goodput-singlechannel}), goodput is obtained by first determining the average number of
transmissions or equivalently the average time required to send one codeword to the destination successfully.
Next result provides the average time in the AF scenario.
\begin{prop} \label{prop:AF}
In AF relaying under the ARQ retransmission scheme with 3 states described in this section, the average time needed to successfully transmit one codeword from the source to the destination is given by
\begin{align}
T_{AF}=\frac{Tp_{1}+2Tp_{2}+2Tp_{3}}{1-p_{3}}\label{eq:T-AF}.
\end{align}
where $p_1, p_2,$ and $p_3$ are given in (\ref{eq:p_1AF}) -- (\ref{eq:p_3AF}), and $T$ is the time needed to transmit one codeword in one attempt over any link.
\end{prop}
\emph{Proof} See Appendix \ref{app:AF}.

The goodput in AF networks is defined as the ratio of
$L$ bits (i.e., the number of information bits in each codeword) to the average time required to successfully send one codeword:
\begin{align}
\eta_{AF}&=\frac{L}{T_{AF}}=\frac{R(1-p_{3})}{p_{1}+2p_{2}+2p_{3}}=\frac{R(1-\varepsilon_{1}\varepsilon_{2})}{1+\varepsilon_{1}}\nonumber\\
&=\frac{R(1-(1-e^{-\frac{2^R-1}{\gamma}})(1-\sqrt{\xi}K_{1}\left(\sqrt{\xi}\right)e^{\frac{-(2^R-1)}{\frac{\gamma}{k^\alpha+(1-k)^\alpha}}}))}{2-e^{-\frac{2^R-1}{\gamma}}}\label{eq:eta-AF}.
\end{align}
where $R$ is the transmission rate in bits/s. The following result identifies the optimal $k$ that maximizes the goodput, and shows that the relay should be located halfway between the source and destination.
\begin{prop}\label{prop:optimal_k}
For any given transmission rate $R$ and SNR $\gamma$, the AF goodput $\eta_{AF}$ is maximized at $ k = 0.5$.
\end{prop}

\emph{Proof:} We first note that
$\sqrt{\xi}K_{1}(\sqrt{\xi})$ is a monotonically
decreasing positive function of $\xi$, and $\xi$ monotonically
increases as $k(1-k)$ increases where $k\in(0,1)$. Hence,
$\sqrt{\xi}K_{1}(\sqrt{\xi})$ is minimized when $k(1-k)$ is
maximized at $k=0.5$. Moreover,
$e^{\frac{-(2^R-1)}{\frac{\gamma}{k^\alpha+(1-k)^\alpha}}}$
monotonically decreases as $k^\alpha+(1-k)^\alpha$ decreases. The function $ f(k)=k^\alpha+(1-k)^\alpha$
with $\alpha>1$ is minimized at $k=0.5$. Therefore,
$e^{\frac{-(2^R-1)}{\frac{\gamma}{k^\alpha+(1-k)^\alpha}}}$ is
minimized at $k=0.5$. Hence, $\sqrt{\xi}K_{1}\left(\sqrt{\xi}\right)e^{\frac{-(2^R-1)}{\frac{\gamma}{k^\alpha+(1-k)^\alpha}}}$
is minimized and $\eta_{AF}$ is maximized at $k=0.5$ regardless of the values of $R$ and $\gamma$. \hfill $\square$

\subsection{Goodput in DF Mode} 
\begin{figure}
\begin{center}
\includegraphics[width = 0.33\textwidth]{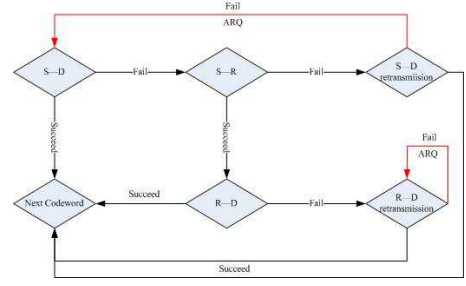}
\caption{Transmission Flow in DF} \label{fig:DF-routing}
\end{center}
\end{figure}

In the DF mode, relay forwards the
packets to the destination only if the S-D link is in outage and S-R
link is successful. So, there are 4 mutually exclusive working
states in the DF mode with the following  ACK/NACK feedback schemes:
1) S-D link is successful and destination sends back ACK to have the
source to send the next codeword; 2) both the S-D and S-R links are
in outage and both send back NACKs to request the  source to
retransmit the codeword; 3) S-D link is in outage, but S-R and R-D
links are successful, and source transmits the next codeword in the
next block after receiving ACK from the destination; 4. S-D link is
in outage, S-R link is successful, but R-D link is in outage. In
this state, destination sends an NACK  but requests retransmissions
from the relay since the relay has successfully decoded the
codeword. Hereby, we see two different types of ARQ retransmissions.
In State 2, source receives NACKs from both relay and destination
and then retransmits, which could be considered as an outer-loop
ARQ. In State 4, relay retransmits to the destination, and  this
could be considered as an inner-loop ARQ. The inner-loop is
beneficial when the relay is closer to the destination than the
source is. The transmission flow can be seen in Fig.
\ref{fig:DF-routing}.

Again we normalize the variance on the S-D link as $\sigma_{s,d}^2=1$ and by
using a generalized distance factor $k=\frac{d_{s,r}}{d_{s,d}},
(k\in(0,1))$, we have the following outage probabilities on the S-D, S-R
and R-D links, respectively,
\begin{align}
\varepsilon_{1}&=1-e^{(-\frac{2^R-1}{\gamma})}\\
\varepsilon_{2}&=1-e^{(-\frac{k^\alpha(2^R-1)}{\gamma})}\\
\varepsilon_{3}&=1-e^{(-\frac{(1-k)^\alpha(2^R-1)}{\gamma})}.
\end{align}
Now, the probabilities of the 4 states are given by
\begin{align}
p_{1}&=1-\varepsilon_{1}\\
p_{2}&=\varepsilon_{1}\varepsilon_{2}\\
p_{3}&=\varepsilon_{1}(1-\varepsilon_{2})(1-\varepsilon_{3})\\
p_{4}&=\varepsilon_{1}(1-\varepsilon_{2})\varepsilon_{3}.
\end{align}

Similarly as in the AF case, we first determine
the average time needed to successfully transmit one
codeword. 
\begin{prop} \label{prop:DF}
In DF relaying under the ARQ retransmission scheme with 4 above-described states,  the average time needed for the successful transmission of one codeword is
\begin{align}
T_{DF}=\frac{Tp_{1}+Tp_{2}+2Tp_{3}+(2+\frac{1}{1-\varepsilon_{3}})Tp_{4}}{1-p_{2}}.
\end{align}
\end{prop}

\emph{Proof}: See Appendix \ref{app:DF}.

Hence, the goodput is
\begin{align} \label{eq:eta-DF}
\eta_{DF}=\frac{L}{T_{DF}}=\frac{R(1-p_{2})}{p_{1}+p_{2}+2p_{3}+(2+\frac{1}{1-\varepsilon_{3}})p_{4}}.
\end{align}

\section{Numerical Results}
In this section, we present the numerical results. In particular, we compute the goodput in AF and DF relaying scenarios by using the expressions in (\ref{eq:eta-AF}) and  (\ref{eq:eta-DF}) and investigate the network performance as the transmission rate $R$, normalized relay location $k$, and SNR
$\gamma$ vary.

\begin{figure}
\begin{center}
\includegraphics[width = 0.4\textwidth]{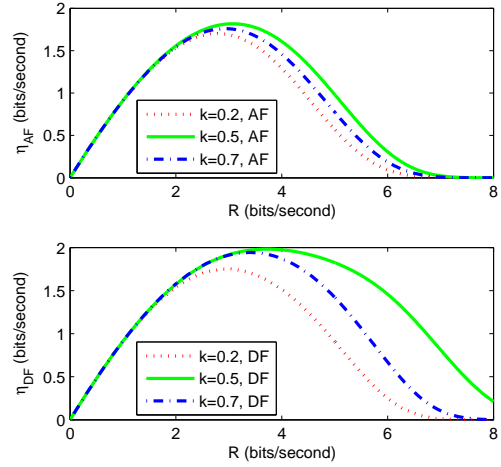}
\caption{Goodput $\eta$ vs $R$} \label{fig:AF/DF-R}
\end{center}
\end{figure}

We first look at $\eta_{AF}$ (goodput in the AF mode) as a function of $R$
with $\gamma=10 dB$ and $\alpha=3.12$ \cite{VE}. The upper subplot
in Fig. \ref{fig:AF/DF-R} clearly shows for any given normalized
relay location $k$, goodput is maximized at a certain unique optimal rate $R^{*}$. At rates  below  $R^{*}$, even though the link
reliabilities are comparatively high, $\eta_{AF}$ is low due
to small $R$. At rates higher than
$R^{*}$, the decrease in link reliability and increased number of retransmissions counteracts the increased
rate, resulting again in lower $\eta_{AF}$. Similar results are also observed in the lower subplot,
which plots the goodput $\eta_{DF}$ in the DF mode  with
respect to transmission rate $R$.

\begin{figure}
\begin{center}
\includegraphics[width = 0.4\textwidth]{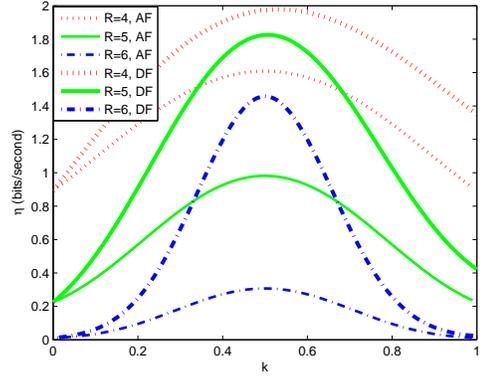}
\caption{Goodput $\eta$ vs Relay Location
$k$}\label{fig:rate-location}
\end{center}
\end{figure}

In Fig.
\ref{fig:rate-location}, we fix the transmission rate $R$ at different values and plot the goodput
$\eta$ for both AF and DF as a function of the normalized relay location $k$. %
Verifying the result of Proposition \ref{prop:optimal_k}, we observe that the AF goodput $\eta_{AF}$ is optimized at $k = 0.5 $ regardless of the value of $R$.
In DF, the optimal relay
location is close to $k=0.5$ for large values of $R$. However, we see that when $R$ is relatively low (i.e., when $R = 4$ bits/s), $k$ is slightly larger than 0.5 meaning that relay should be placed farther away from the source and closer to the destination. Note also that a higher goodput is achieved when $R = 4$ bits/s.  Finally, we observe that at the optimal relay location, DF  relaying provides higher goodput than AF relaying for the same transmission rate.

\begin{figure}
\begin{center}
\includegraphics[width = 0.4\textwidth]{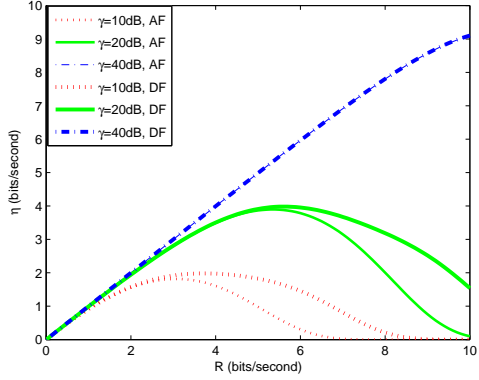}
\caption{Goodput $\eta$ vs $R$ at Different SNR}\label{fig:SNR-rate}
\end{center}
\end{figure}

In Fig. \ref{fig:SNR-rate}, we plot $\eta$ as a function of $R$ at
different SNR values in both AF and DF when the generalized relay
location is set at $k = 0.5$. When SNR is maintained at a moderate level,
DF outperforms AF in terms of providing higher goodput at any transmission
rate. At very high SNR, AF and DF modes have almost identical
performance curves. Since the link reliability is so high that very
few ARQ rounds are needed to achieve the successful packet delivery,
the goodput is actually approaching the transmission rate $R$.
Secondly, in either mode, we see as $\gamma$ increases, the optimal
goodput $\eta^*$ and $R^*$ which maximizes $\eta$ are both
increasing. 

\begin{figure}
\begin{center}
\includegraphics[width = 0.4\textwidth]{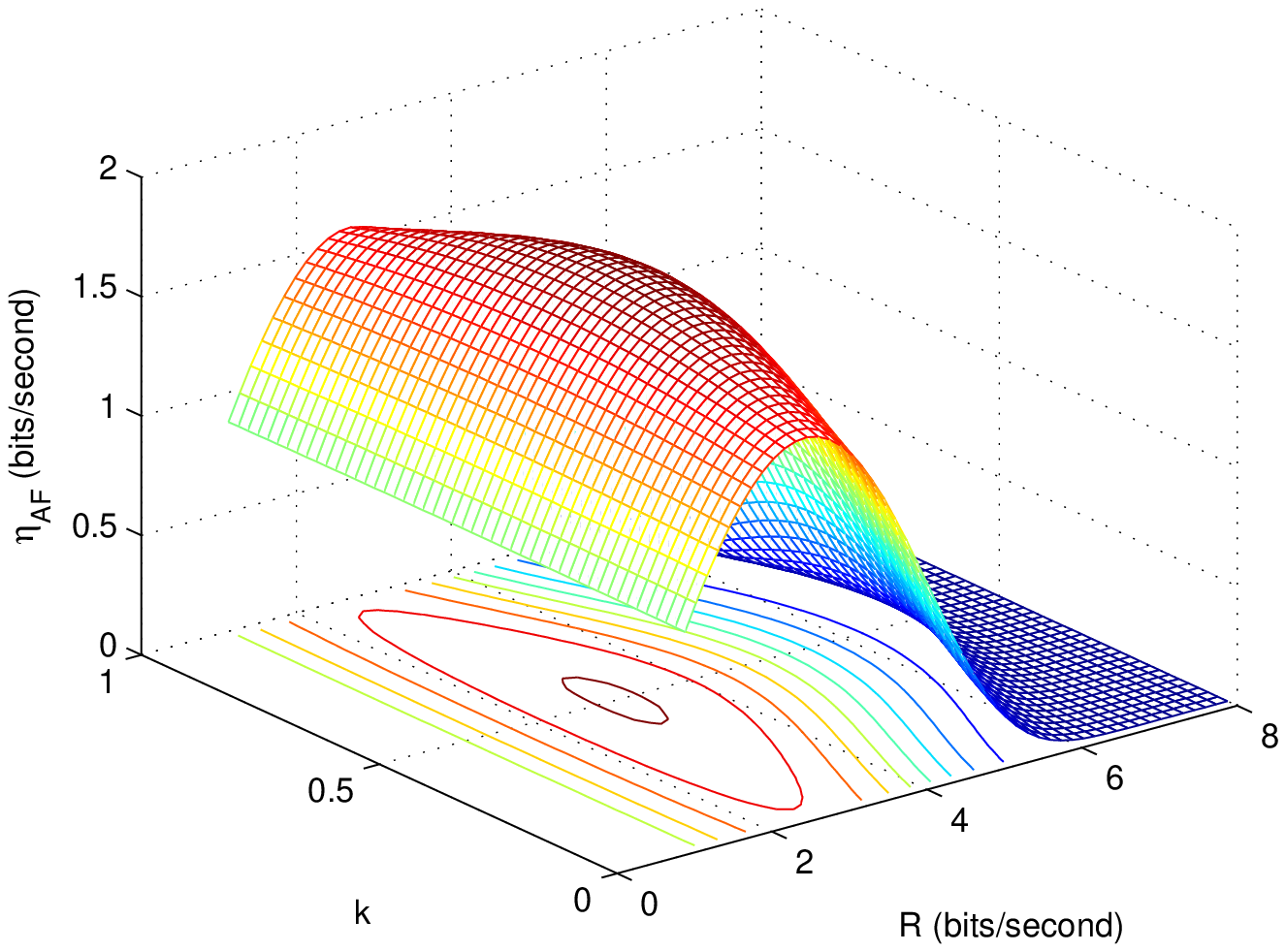}
\caption{Goodput vs $k$ and $R$ in AF} \label{fig:AF-Location-R}
\end{center}
\end{figure}

\begin{figure}
\begin{center}
\includegraphics[width = 0.4\textwidth]{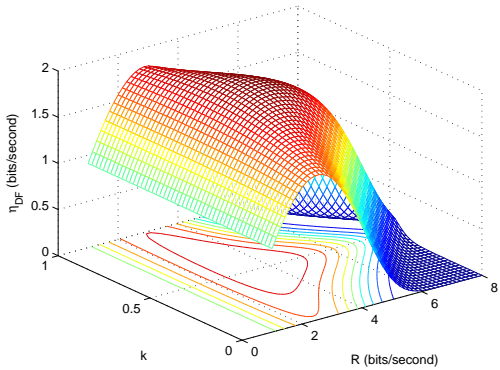}
\caption{Goodput vs $k$ and $R$ in DF} \label{fig:DF-Location-R}
\end{center}
\end{figure}
In Fig. \ref{fig:AF-Location-R} and Fig. \ref{fig:DF-Location-R}, the goodput is plotted as both
the relay location and transmission rate are varied. It is assumed that SNR $\gamma=10 dB$. As is shown in the contours, the
goodput surface isn't ideally concave, but there is always a pair of
optimal relay location $k^*$ and transmission rate $R^*$ that
maximizes the goodput.

\begin{figure}
\begin{center}
\includegraphics[width = 0.4\textwidth]{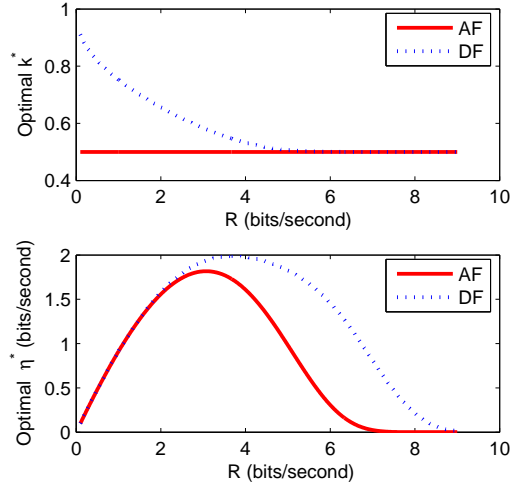}
\caption{Optimal Relay Location and Goodput vs $R$}\label
{fig:Location-Goodput}

\end{center}
\end{figure}

In Fig. \ref{fig:Location-Goodput} , we assume $\gamma=10 dB$ and
plot the optimal $k*$ and the optimal goodput $\eta^*$ as a function
of the transmission rate $R$ in both AF and DF. The upper subplot
confirms again the result in Proposition \ref{prop:optimal_k} and
shows that $k^* = 0.5$ in AF. In the DF mode, we see that $k^*$
decreases from 1 and approaches 0.5 asymptotically as $R$ increases,
which is in compliance with our observation in Fig.
\ref{fig:rate-location}. In the lower subplot, the optimal goodput
$\eta^*$ achieved at the optimal $k^*$ is shown, where we see that
DF outperforms AF and provides higher goodput for any given
transmission rate $R$.

%
%

\section{Conclusion}
In this paper, we have performed a cross-layer optimization to maximize
the goodput of cooperative networks that employ ARQ schemes and work in either AF or DF
relaying mode. For both AF and DF schemes, we have described the retransmission mechanisms and determined the average time required to send the information correctly from the source to the destination. Subsequently, we have formulated the goodput achieved in both relaying modes. We investigated the impact of transmission rates and relay location on the goodput. We have shown that having the relay halfway between the source and destination maximizes the goodput. Through numerical results, we have identified the optimal transmission rates that maximize the goodput. We have also numerically analyzed the optimal relay location in DF. We have shown that DF generally provides superior performance.

\appendix
\subsection {Proof of Proposition \ref{prop:AF}} \label{app:AF}
\begin{figure}
\begin{center}
\includegraphics[width = 0.3\textwidth]{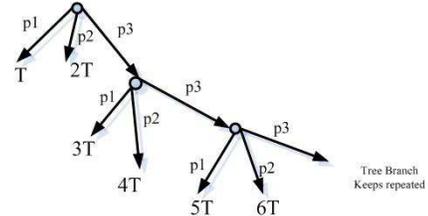}
\caption{Transmission Tree Structure in AF}\label {fig:AF-Tree}
\end{center}
\end{figure}

In the AF mode, we have three networks states and the ARQ is
triggered when both S-D and S-R-D links fail. Therefore,
the transmission flow  can be illustrated as a tree structure as depicted in
Fig.  \ref{fig:AF-Tree} as the ARQ round progresses.
Each branch corresponds to a time consumption with a certain
probability. We model the time required to
successfully transmit one codeword as a random variable and denote it as $X$. It can be easily seen that $X$
has the following distribution by
\begin{align}
X= \left\{\begin{array}{ll}T,&\text{$p_{1}$}\\
2T,&\text{$p_{2}$}\\
3T,&\text{$p_{3}p_{1}$}\\
4T,&\text{$p_{3}p_{2}$}\\
5T,&\text{$p_{3}^2p_{1}$}\\
6T,&\text{$p_{3}^2p_{2}$}\\\vdots
\end{array}\right.
\end{align}
For odd number of time slots, the probability distribution is
\begin{align}
p((2k+1)T)= \begin{array}{ll}p_{3}^kp_{1},&\text{ for
$k=0,1,2\cdots$}\\
\end{array}
\end{align}
For even number of time slots, the probability distribution is
\begin{align}
p((2k+2)T)= \begin{array}{ll}p_{3}^kp_{2},&\text{ for
$k=0,1,2\cdots$}\\
\end{array}
\end{align}
So, the average time $E(X) = T_{AF}$ is
\begin{align}
T_{AF}&=\sum_{k=0}^{\infty}(2k+1)Tp_{3}^kp_{1}+\sum_{k=0}^{\infty}(2k+2)Tp_{3}^kp_{2}\nonumber\\
&=T\left[p_{1}(\frac{2p_{3}}{(1-p_{3})^2}+\frac{1}{1-p_{3}})+2p_{2}(\frac{p_{3}}{(1-p_{3})^2}+\frac{1}{1-p_{3}})\right]\nonumber\\
&=\frac{(p_{1}+2p_{2}+2p_{3})T}{1-p_{3}}.
\end{align}
 \hfill $\square$

\subsection {Proof of Proposition \ref{prop:DF}} \label{app:DF}
\begin{figure}
\begin{center}
\includegraphics[width = 0.35\textwidth]{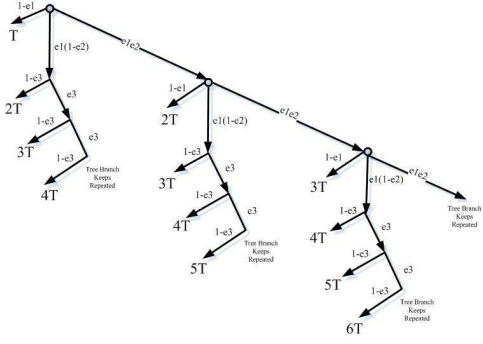}
\caption{Transmission Tree Structure in DF}\label {fig:DF-Tree}
\end{center}
\end{figure}

In the DF mode, the transmission flow  can be illustrated as in Fig.
\ref{fig:DF-Tree}, where the long branches occurring with probability $\varepsilon_{1}\varepsilon_{2}$ and extending to the right are generated when the
outer ARQ round is triggered. Recall that the outer ARQ is triggered when both S-R and S-D links fail, requiring the source to retransmit. Note that there is also a downward progression starting initially with probability $\varepsilon_{1}(1-\varepsilon_{2})$. This progression occurs when the inner ARQ rounds are triggered, in which relay, having received the packets successfully from the source, retransmits when the R-D link fails.
In DF relaying, the random variable $X$, which is time required to one codeword successfully, can be expressed as
\begin{align}
X= \left\{\begin{array}{ll}T,&\text{$1-e_{1}$}\\
Y_{1},&\text{$\varepsilon_{1}(1-\varepsilon_{2})$}\\
Z_{1},&\text{$\varepsilon_{1}\varepsilon_{2}$}
\end{array}\right.
\end{align}
where $Y_{1}$ and $Z_{1}$ are also random variables whose
distributions are as follows:
\begin{align}
Y_{1}= \left\{\begin{array}{ll}2T,&\text{$1-\varepsilon_{3}$}\\
3T,&\text{$\varepsilon_{3}(1-\varepsilon_{3})$}\\
\vdots\\
nT,&\text{$\varepsilon_{3}^{n-2}(1-\varepsilon_{3})$}
\end{array}\right.,
\quad
Z_{1}= \left\{\begin{array}{ll}2T,&\text{$1-\varepsilon_{1}$}\\
Y_{2},&\text{$\varepsilon_{1}(1-\varepsilon_{2})$}\\
Z_{2},&\text{$\varepsilon_{1}\varepsilon_{2}$}
\end{array}\right..
\end{align}
Above, we have defined
\vspace{-.2cm}
\begin{align}
Y_{2}= \left\{\begin{array}{ll}3T,&\text{$1-\varepsilon_{3}$}\\
4T,&\text{$\varepsilon_{3}(1-\varepsilon_{3})$}\\
\vdots\\
nT,&\text{$\varepsilon_{3}^{n-3}(1-\varepsilon_{3})$}
\end{array}\right..
\end{align}
$Z_2$ can be defined similarly as $Z_1$ but now in terms of $Y_3$ and $Z_3$. Repeating this procedure, we have random variables defined in a nested fashion with a certain pattern.
By examining the characteristics of $Y_{k}$, we can easily find its mean value as
\begin{align}
E(Y_{k})=(1-\varepsilon_{3})T\left(\frac{\varepsilon_{3}}{(1-\varepsilon_{3})^2}+\frac{k+1}{1-\varepsilon_{3}}\right).
\end{align}
Now, the expected value of $X$ is
\begin{align}
E(X)=&(1-\varepsilon_{1})T+\varepsilon_{1}(1-\varepsilon_{2})E(Y_{1})+\varepsilon_{1}\varepsilon_{2}E(Z_{1})\nonumber\\
=&(1-\varepsilon_{1})T+\varepsilon_{1}(1-\varepsilon_{2})E(Y_{1})+\varepsilon_{1}\varepsilon_{2}\cdot\nonumber\\
&[(1-\varepsilon_{1})2T+\varepsilon_{1}(1-\varepsilon_{2})E(Y_{2})+\varepsilon_{1}\varepsilon_{2}\cdot\nonumber\\
&\left((1-\varepsilon_{1})3T+\varepsilon_{1}(1-\varepsilon_{2})E(Y_{3})+\varepsilon_{1}\varepsilon_{2}E(Z_{3})\right)]\nonumber\\
=&\sum_{k=0}^{\infty}(\varepsilon_{1}\varepsilon_{2})^k(1-\varepsilon_{1})(k+1)T+\sum_{k=0}^{\infty}(\varepsilon_{1}\varepsilon_{2})^k\varepsilon_{1}(1-\varepsilon_{2})E(Y_{k+1})\nonumber\\
=&\frac{T(1-\varepsilon_{1}\varepsilon_{2})(1+\varepsilon_{1}-\varepsilon_{3}-\varepsilon_{1}\varepsilon_{2})}{(1-\varepsilon_{1}\varepsilon_{2})^2(1-\varepsilon_{3})}\nonumber\\
=&\frac{Tp_{1}+Tp_{2}+2Tp_{3}+(2+\frac{1}{1-\varepsilon_{3}})Tp_{4}}{1-p_{2}}.
\end{align}
 \hfill $\square$

\end{document}